\documentclass[twocolumn]{aastex701}
	
\usepackage{color}
\usepackage{amsmath}
\usepackage{multirow}

\shorttitle{AGN Large-scale Environments Revealed by JWST}
\shortauthors{Jiang et al.}

\begin{document}
\title{Large-scale Environmental Dependence of AGN Activity Revealed by JWST Pa$\alpha$ Emitters}

\author[0009-0003-6747-2221]{Danyang Jiang}
\email{jiangdy@stu.pku.edu.cn}
\affiliation{Department of Astronomy, School of Physics, Peking University, Beijing 100871, China}
\affiliation{Kavli Institute for Astronomy and Astrophysics, Peking University, Beijing 100871, China}

\author[0000-0003-4176-6486]{Linhua Jiang}
\email{jiangkiaa@pku.edu.cn}
\affiliation{Department of Astronomy, School of Physics, Peking University, Beijing 100871, China}
\affiliation{Kavli Institute for Astronomy and Astrophysics, Peking University, Beijing 100871, China}

\author[0000-0003-0964-7188]{Shuqi Fu}
\affiliation{Department of Astronomy, School of Physics, Peking University, Beijing 100871, China}
\affiliation{Kavli Institute for Astronomy and Astrophysics, Peking University, Beijing 100871, China}
\affiliation{Jodrell Bank Centre for Astrophysics, University of Manchester, Oxford Road, Manchester UK}
\email{fushuqi@stu.pku.edu.cn}

\author[0000-0002-2420-5022]{Zijian Zhang}
\affiliation{Department of Astronomy, School of Physics, Peking University, Beijing 100871, China}
\affiliation{Kavli Institute for Astronomy and Astrophysics, Peking University, Beijing 100871, China}
\email{zjz.kiaa@stu.pku.edu.cn}

\author[0000-0002-4765-1500]{Jie Chen}
\affiliation{Department of Astronomy, School of Physics, Peking University, Beijing 100871, China}
\affiliation{Kavli Institute for Astronomy and Astrophysics, Peking University, Beijing 100871, China}
\email{jiechen@stu.pku.edu.cn}

\author[0009-0003-2490-4434]{Chenxi Bao}
\affiliation{Department of Astronomy, School of Physics, Peking University, Beijing 100871, China}
\email{2300011608@stu.pku.edu.cn}

\author[0009-0004-6535-9038]{Yihe Wang}
\affiliation{Department of Astronomy, School of Physics, Peking University, Beijing 100871, China}
\email{wangyidaihe@stu.pku.edu.cn}

\author[0009-0007-0864-7094]{Junya Arita}
\email{jarita@astron.s.u-tokyo.ac.jp}
\affiliation{Department of Astronomy, School of Science, The University of Tokyo, 7-3-1, Hongo, Bunkyo, Tokyo, 113-0033, Japan}

\author[]{Darshan Kakkad}
\email{darshankakkad@gmail.com}
\affiliation{Centre for Astrophysics Research, Department of Physics, Astronomy and Mathematics, University of Hertfordshire, Hatfield, AL10 9AB, UK}

\author[0000-0002-4544-8242]{Jan-Torge Schindler}
\email{jtschindler@hs.uni-hamburg.de}
\affiliation{Hamburg Observatory, University of Hamburg, Gojenbergsweg 112, D-21029 Hamburg, Germany}

\begin{abstract}
The role of the large-scale environment in regulating AGN activity remains an open question. We investigate the Mpc-scale environments of 70 AGNs at $z=1.1$–$1.5$ using a highly complete sample of 3,419 Pa$\alpha$ emitters with accurate spectroscopic redshifts from JWST slitless spectroscopic survey. This dataset allows us to robustly trace the environments and quantify the environmental dependence of AGN properties. We measure the eighth-nearest-neighbor surface density of Pa$\alpha$ emitters around each AGN, probing $\sim$1.5 pMpc scale environments.
We find that this environmental density exhibits a weak positive correlation with radio luminosity. Radio-bright AGNs tend to reside in statistically denser environments than their radio-faint counterparts. The environmental density also has a positive correlation with the Eddington ratio and a weak negative correlation with the central black hole mass, implying more efficient black hole accretion in dense environments.
We construct Pa$\alpha$ luminosity functions (LFs) in AGN environments and find that they are systematically elevated relative to the field Pa$\alpha$ LF. In addition, their faint-end slopes are shallower, suggesting accelerated galaxy growth and enhanced quenching of star formation. As Pa$\alpha$ emitters trace gas-rich, star-forming galaxies, our results support a scenario in which the large-scale cold gas promotes AGN activity at cosmic noon.
\end{abstract}


\graphicspath{{./}{figures/}}

\section{Introduction} \label{sec:introduction}

Active galactic nuclei (AGNs) powered by mass accretion onto supermassive black holes (SMBHs) represent a key phase in the evolutionary history of galaxies. Tight correlations between SMBHs and host-galaxy properties at low redshift suggest a co-evolution of black-hole growth and galaxy assembly \citep[e.g.,][]{gebhardt2000,ferrarese2000,marconi2003,kormendy2013}. Beyond the host galaxy, the large-scale environment may also influence the triggering and evolution of AGN activity. Dense environments can facilitate SMBH fueling through several physical processes such as efficient cold-gas transport, enhanced galaxy mergers, and frequent tidal interactions \citep[e.g.,][]{silverman2011,liu2012,gaspari2013,goulding2018,storchi-bergmann2019}.  
Characterizing the environmental dependence of AGNs is essential for understanding the role of large-scale environments in SMBH growth and galaxy evolution.

Over the past two decades, extensive efforts have been devoted to investigating the environments of AGNs. Among the various AGN populations, radio AGNs are found to preferentially reside in overdense environments. Radio AGNs exhibit stronger clustering and are typically associated with more massive dark matter halos than the radio-quiet galaxies \citep[e.g.,][]{peacock1991,magliocchetti2004,magliocchetti2017,hale2018,croston2019}.
Galaxy searches surrounding radio AGNs further reveal systematically higher environmental densities than galaxies with similar stellar mass ($M_{\mathrm{stellar}}$) and star formation rate (SFR) \citep[e.g.,][]{best2004a,bardelli2010,wylezalek2013,hatch2014,malavasi2015,magliocchetti2022a}. Despite this overdensity, the role of environment in regulating AGN activity remains unclear.

Different studies report inconsistent results of the AGN large-scale environmental dependence on AGN and host-galaxy properties, from the nearby Universe to cosmic noon. The relation between radio power and environmental density is controversial. While some studies have reported no significant dependence \citep[e.g.,][]{wylezalek2013,kolwa2019}, others have found a negative trend \citep[e.g.,][]{malavasi2015} or a positive trend \citep[e.g.,][]{falder2010}. The environmental dependence of $M_{\mathrm{BH}}$ and accretion rate shows no significant trends with substantial scatter \citep[e.g.,][]{hatch2014,yang2018}.
Such studies mostly rely on photometric data and ground-based spectroscopic surveys. The former suffer from photometric-redshift uncertainties, while the latter are restricted to bright galaxies and provide an incomplete census of the surrounding galaxy population.

Emission-line galaxies, such as Ly$\alpha$, H$\alpha$, and [O III] emitters, provide flux-complete galaxy samples with accurate redshifts for environmental studies from $z\sim1$ to the early Universe. 
Narrow-band-selected emission-line galaxies have been widely used to investigate AGN environments \citep[e.g.,][]{matsuda2011,mazzucchelli2017b,garcia-vergara2019a,wang2024d,lambert2024,arita2026a}. Slitless spectroscopy by the Hubble Space Telescope (HST) and the James Webb Space Telescope (JWST) has also been used to probe galaxies surrounding luminous quasars \citep{stott2020,wang2026}. Owing to the small sky coverage and wavelength range of those surveys, these kinds of studies focus on individual AGNs or relatively small samples with $\la 25$ AGNs. This limits the statistical characterization of environmental trends representative of the broader AGN population.

Recently, JWST NIRCam wide-field slitless spectroscopy (WFSS) enables deep spectroscopic surveys over large contiguous cosmological volumes \citep[e.g.,][]{meyer2024,lin2025a,sun2025}, opening a new window for AGN environmental studies. In particular, COSMOS-3D (GO\#5893; PI: K. Kakiichi) is a large Cycle 3 NIRCam+MIRI imaging and F444W NIRCam WFSS survey covering $0.33~\mathrm{deg}^2$ within the COSMOS field, which facilitates several studies \citep[e.g.,][]{lin2026c,meyer2026,jiang2026}. 
Based on the WFSS and multi-band imaging data, \cite{fu2026b} constructed a large and homogeneous catalog of 12,275 emission-line galaxies in the COSMOS-3D field.
Among the identified emission-line galaxies, Pa$\alpha$ emitters at cosmic noon constitute the largest fraction due to the elevated star formation rate density. 
As a near-infrared (NIR) hydrogen recombination line, Pa$\alpha$ is a robust tracer of star formation, barely affected by dust attenuation.
Therefore, Pa$\alpha$ emitters are particularly valuable for environmental investigations. 

In this study, we investigate AGN environments based on a highly complete Pa$\alpha$ emitter sample at $z=1.1-1.5$ in the COSMOS-3D field. This dataset enables the first statistical study of AGN environmental dependence based on a large spectroscopic emission-line galaxy sample.
The structure of this Letter is as follows. We introduce our datasets and the sample construction in Section \ref{sec:data}. We describe our environmental density estimation and show the results in Section \ref{sec:results}. We summarize the Letter and provide implications for the scenario of large-scale AGN environments in Section \ref{sec:sum}. We adopt a standard $\Lambda$CDM cosmology with $H_0$=70 km s$^{-1}$ Mpc$^{-1}$, $\Omega_\mathrm{m}$=0.3, and $\Omega_\mathrm{\Lambda}$=0.7. All the magnitudes are reported in the AB system \citep{Oke1983}.

\section{Data and Sample Construction} \label{sec:data}

\subsection{Highly Complete Pa\texorpdfstring{$\alpha$}{α} Emitter Sample} \label{sec:emitters}

We built our parent sample of Pa$\alpha$ emitters based on the comprehensive emission-line galaxy catalog provided by \cite{fu2026b}. They reduced the F444W NIRCam WFSS data in the COSMOS-3D field, extracted grism spectra for all the photometric sources, and detected emission lines in both 1D and 2D spectra. 
The $3\sigma$ line-detection depth corresponds to an F444W-equivalent magnitude of $m_{\rm F444W}\simeq27.5$ mag, which is brighter than the F444W $5\sigma$ depth of $\sim28$ mag \citep{shuntov2025b}. The photometry is therefore sufficiently deep for the emission-line catalog with negligible incompleteness.
To remove contamination and identify emission lines, they adopted a multi-step framework that simultaneously considers photometric redshifts, spectral energy distribution (SED) shapes, the locations of multiple emission lines, line morphologies, and available spectroscopic redshifts. The resulting catalog contains 5,190 Pa$\alpha$ emitters at $1.07 < z < 1.66$, with $M_{\mathrm{stellar}}$ and SFR derived from SED fitting provided for each source \cite{fu2026b}.

To robustly measure environments using Pa$\alpha$ emitters, we further evaluated the sample completeness. The incompleteness mainly arises from the wavelength coverage of the grism observations and the line detections. We performed simulations to derive the Pa$\alpha$ selection completeness as a function of line flux and redshift (S. Fu et al. in preparation; see also \citealt{fu2025f}). The simulation procedure and resultant selection function is presented in Appendix~\ref{sec:appendixA}. 

\begin{figure*}[t]
\epsscale{1.2}
\plotone{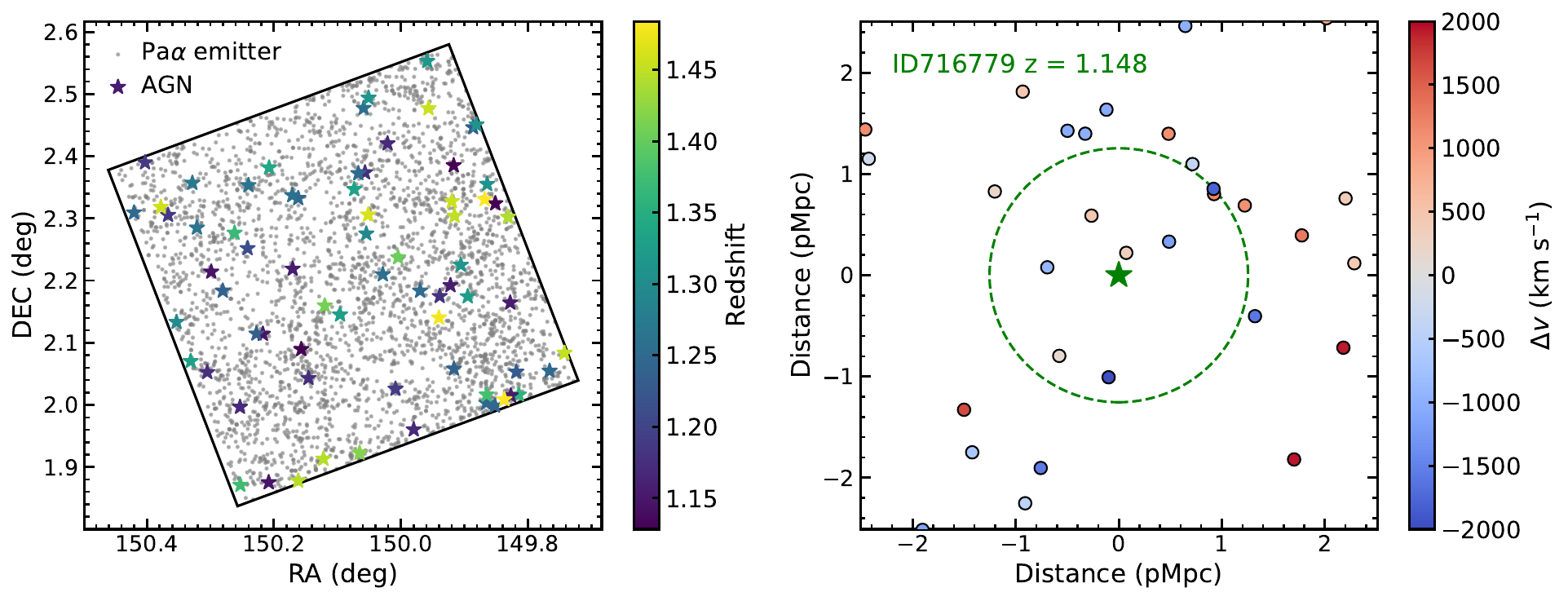}
\caption{Distribution of the Pa$\alpha$ emitters and AGNs. The left panel shows the sky distribution of the Pa$\alpha$ emitters (gray dots) and AGNs (stars color-coded by redshift) in the COSMOS-3D field, with the survey footprint outlined in black. The right panel presents the Mpc-scale environment around an AGN (green star). Pa$\alpha$ emitters within $\Delta v \leq \pm 2000$ km s$^{-1}$ relative to the central AGN are shown as circles color-coded by $\Delta v$, with the green dashed circle marking the eighth-nearest neighbor region.
 \label{fig:distri}}
\end{figure*}

We restricted the Pa$\alpha$ emitter sample to a high-completeness subsample, since completeness corrections become increasingly uncertain at the low-completeness end. 
We first limited the redshift range to $1.1 < z < 1.5$ to avoid ineffective wavelength regions of the grism spectroscopy. We then applied a line-flux threshold of 0.035 mJy, corresponding to the 5$\sigma$ detection and $5.4 \times 10^{-18}$ erg s$^{-1}$ cm$^{-2}$ at observed-frame 4.4 $\mu$m. With these selections, the average completeness in each simulated flux bin exceeds 60\%. The resulting parent sample consists of 3,419 Pa$\alpha$ emitters, of which 74\% have completeness values above 80\%. In the subsequent environmental density measurement, the incompleteness of each Pa$\alpha$ emitter will be further corrected. We show the sky distribution of this Pa$\alpha$ emitter sample in the left panel of Figure \ref{fig:distri}.

\subsection{AGN Sample} \label{sec:paa_agns}

We constructed a Pa$\alpha$ AGN sample consisting of broad-line and X-ray AGNs. 
We first adopted 37 broad-line Pa$\alpha$ AGNs at $1.1<z<1.5$ identified by \cite{jiang2026}, which also provide robust measurements of the intrinsic AGN properties including Pa$\alpha$-based $M_{\rm BH}$, bolometric luminosities ($L_{\rm bol}$), and $\lambda_{\rm Edd}$.
To supplement the AGN sample, we then cross-matched the Pa$\alpha$ emitter catalog with the Chandra COSMOS Legacy Survey catalog \citep{civano2016} and selected sources with X-ray luminosities $L_{\rm 0.5-10keV}>3\times10^{42}$ erg s$^{-1}$. We further excluded two X-ray sources, the special merging system known as the Cosmic Owl \citep{li2026c} and one AGN located too close to our redshift boundary for the environment measurement. 
Among the 37 broad-line Pa$\alpha$ AGNs, 30 also satisfy the X-ray selection.
The final Pa$\alpha$ AGN sample contains 70 identified AGNs, enabling a statistical investigation of environmental dependence.

We cross-matched the Pa$\alpha$ AGN sample with the Very Large Array (VLA) radio catalogs \citep{smolcic2017,jin2018}. Among the 70 AGNs, 46 have radio detections with $\geq3\sigma$ significance in the deep 3 GHz observations. We calculated the radio spectral index $\alpha$ (where $f_\nu\propto\nu^\alpha$) for 25 AGNs detected at both 1.4 and 3 GHz, obtaining a mean value of $\alpha=-0.9$. We adopted this value for the entire sample. We calculated the rest-frame radio fluxes and luminosities for all radio-detected AGNs using the 3 GHz observations.
We also quantified the radio loudness $R=f_{\rm 5GHz}/f_{4400\text{\AA}}$, where $f_{\rm 5GHz}$ is the rest-frame 5 GHz flux and $f_{4400\text{\AA}}$ is the intrinsic AGN flux at rest-frame 4400 \AA~\citep{kellermann1989}. For the broad-line AGNs, \cite{jiang2026} measure the intrinsic AGN emission through JWST image decomposition and SED fitting. They removed the host-galaxy emission and corrected dust attenuation for the AGN component. We performed the same analysis for the additional X-ray AGNs to derive their intrinsic AGN fluxes and calculated the radio loudness.

The observed radio emission may originate from both AGN activity and star formation in the host galaxy. To assess the dominant origin of the radio emission, we calculated the IR-to-radio luminosity ratio $q_{\rm TIR}$, defined as
\begin{equation}
q_{\rm TIR}=\log \left(\frac{L_{\rm TIR}[{\rm W}]}{3.75\times10^{12}[{\rm Hz}]}\right)-\log \left(L_{1.4{\rm GHz}}[{\rm W\ Hz^{-1}}]\right),
\end{equation}
where $L_{\rm TIR}$ is the total IR luminosity integrated over 8--1000 $\mu$m. Star-forming galaxies follow an IR--radio correlation, whereas AGN activity can produce excess radio emission and therefore systematically lower values of $q_{\rm TIR}$ \citep[e.g.,][]{molnar2018,cook2024}.
We cross-matched our AGN sample with the far-IR to (sub-)millimeter photometric catalog from \cite{jin2018} and estimated $L_{\rm TIR}$ by interpolating the observed fluxes. Reliable $q_{\rm TIR}$ measurements were obtained for 45 AGNs, spanning a range of $0.9<q_{\rm TIR}<2.9$. We considered AGNs with $q_{\rm TIR}<2$ as AGN-dominated radio sources \citep{cook2024}.
Owing to the intrinsic scatter of the IR--radio correlation and the coexistence of AGN and star-formation radio emission, $q_{\rm TIR}$ serves primarily as a statistical indicator instead of a diagnostic for individual sources. 

To study the overdense environments traced by radio AGNs, we divided our Pa$\alpha$ AGN sample into the radio-bright and radio-faint AGN samples using a threshold of $\log L_{1.4{\rm GHz}}({\rm W\ Hz^{-1})} = 23.5$. 
This threshold is adopted to split the AGN sample into two similarly sized groups, each containing 35 Pa$\alpha$ AGNs, while ensuring that the radio-bright AGNs have robust radio detections.
We will perform $M_{\mathrm{stellar}}$ and SFR controlled analyses for both samples, respectively.
We show the sky distribution of the Pa$\alpha$ AGN sample in the left panel of Figure \ref{fig:distri}.

\subsection{Control Samples} \label{sec:paa_controls}

We constructed control samples to evaluate the AGN-related environmental effect without biases caused by the host galaxies. The $M_{\mathrm{stellar}}$ and SFR measurements were adopted from \cite{fu2026b}. 
Our Pa$\alpha$ AGNs are hosted by more massive galaxies with median $\log(M_{\mathrm{stellar}}/M_{\odot}) \sim 10.84$, compared to the overall Pa$\alpha$ emitters with median $\log(M_{\mathrm{stellar}}/M_{\odot}) \sim 9.96$.  
At $z\ga1$, enhanced star formation is also observed in galaxies hosting radio AGNs \citep[e.g.,][]{drouart2014,magliocchetti2016,kalfountzou2017}.
We generated control samples for the full AGN sample, the radio-bright AGN sample, and the radio-faint AGN sample, separately.
For each AGN in a given sample, we randomly selected a Pa$\alpha$ emitter with $M_{\mathrm{stellar}}$ within $\pm$0.25 dex and SFR within $\pm$0.5 dex of the AGN host galaxy. These matching tolerances generally represent the smallest 16th–84th percentile ranges of $M_{\mathrm{stellar}}$ and SFR distributions of three AGN samples.
We excluded 435 galaxies with radio detections in this control construction \citep{jin2018}.
We repeated this procedure for 100 times and obtained 100 independent control samples with the same size of the corresponding AGN sample.

\section{Results} \label{sec:results}

\subsection{Surface Density Estimation} \label{sec:density}

To quantify the environmental density, we measure the $N$th-nearest neighbor surface density $\Sigma_{N}$, defined as
\begin{equation}
\Sigma_{N}=\frac{N}{\pi d_{N}^2},
\end{equation}
where $d_{N}$ is the projected distance to the $N$th-nearest neighbor \citep[e.g.,][]{casertano1985,mateus2004}. To reduce foreground and background contamination, we apply a line-of-sight velocity cut of $\Delta v \leq \pm 2000~\mathrm{km\ s^{-1}}$ relative to the central AGN, corresponding to $\Delta z \leq \pm0.015$ at $z=1.3$. Such a velocity window has been commonly adopted in studies of galaxy environments to account for the peculiar motions of galaxies within massive halos and minimize the loss of nearby galaxies \citep[e.g.,][]{lee2019,santana-silva2020}. In addition, this choice can also accommodate the $\sim 300~\mathrm{km\ s^{-1}}$ redshift uncertainty of the grism spectra, which arises from the extended morphology of the Pa$\alpha$ emission.

For each AGN, we identify the eighth-nearest Pa$\alpha$ emitters and compute the eighth-nearest neighbor surface density $\Sigma_8$ with effective corrections. 
We first correct for the selection incompleteness of each emitter. The effective number of emitters is calculated as $N_{\mathrm{eff}} = \sum_{i=1}^{8} 1/p_i(z_i,\mathrm{flux}_i)$, where $p_i(z_i,\mathrm{flux}_i)$ is the selection completeness of the $i$-th emitter.
We then correct the effective area of each AGN environment, accounting for regions falling outside the survey footprint and those masked to avoid bright stars during the photometric source detection \citep{shuntov2025b}. We further exclude two AGNs located at the corners of the survey field for which $d_{8}$ exceeds 8 cMpc. The resulting density measurements probe large-scale AGN environments with a mean physical scale of 1.5 pMpc and a standard deviation of 0.6 pMpc. We present an example of the AGN environment in the right panel of Figure \ref{fig:distri}. The same environmental density measurements were also applied to the control samples.

\subsection{Distributions of Controlled Density} \label{sec:envir}

\begin{figure}[t]
\epsscale{1.2}
\plotone{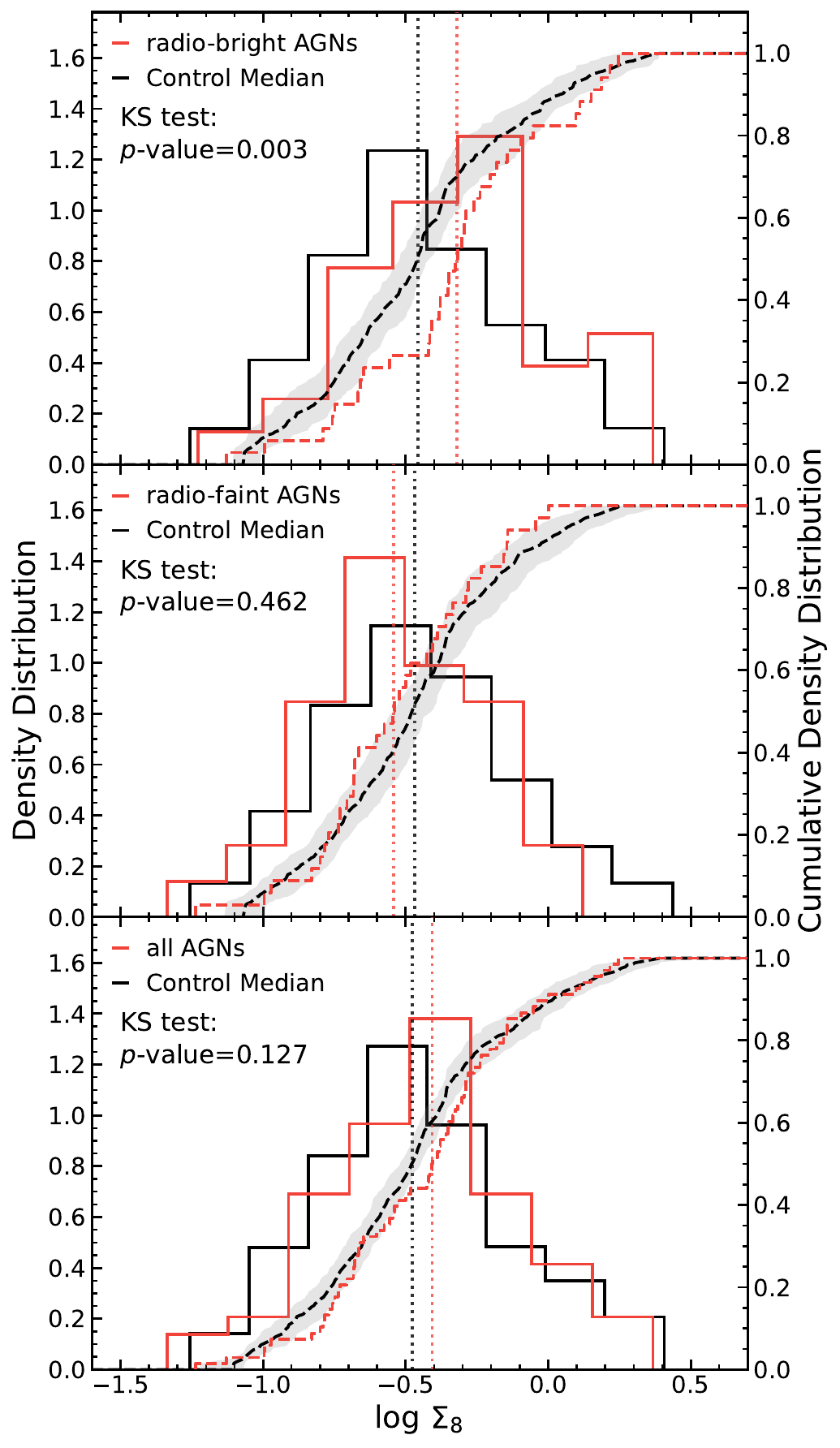}
\caption{Surface density distributions of the AGN and control samples. The top, middle, and bottom panels show the radio-bright, radio-faint, and full AGN samples, respectively. In each panel, the red histogram and dashed curve show the probability density and cumulative distribution of the logarithmic $\Sigma_8$ for the AGN sample. The black histogram and dashed curve show the corresponding distributions for the median control sample. The grey shaded region indicates the $1\sigma$ scatter derived from 100 control samples. The red and black vertical dotted lines mark the median $\Sigma_8$ values of the AGN and control samples, respectively.
 \label{fig:control}}
\end{figure}

We compare the environmental density distributions of the radio-bright, radio-faint, and full AGN samples with those of their corresponding control samples in Figure~\ref{fig:control}. We find that radio-bright AGNs reside in significantly denser environments than the AGNs in the control sample, with a median density excess of 0.14 dex and a Kolmogorov–Smirnov (KS) test $p$-value = 0.003. This result is consistent with previous studies showing that radio AGNs preferentially inhabit overdense environments \citep[e.g.,][]{wylezalek2013,hatch2014,malavasi2015}. Meanwhile, the environmental density distributions of the radio-faint and full AGN samples are indistinguishable from those of their respective control samples. The elevated environmental densities of radio-bright AGNs suggest that dense large-scale environments may enhance AGN radio emission. We further explore the environmental dependence of AGN radio properties in Section~\ref{sec:dependence}.

\subsection{Environmental Dependence of AGN Properties} \label{sec:dependence}

\begin{figure}[t]
\epsscale{1.2}
\plotone{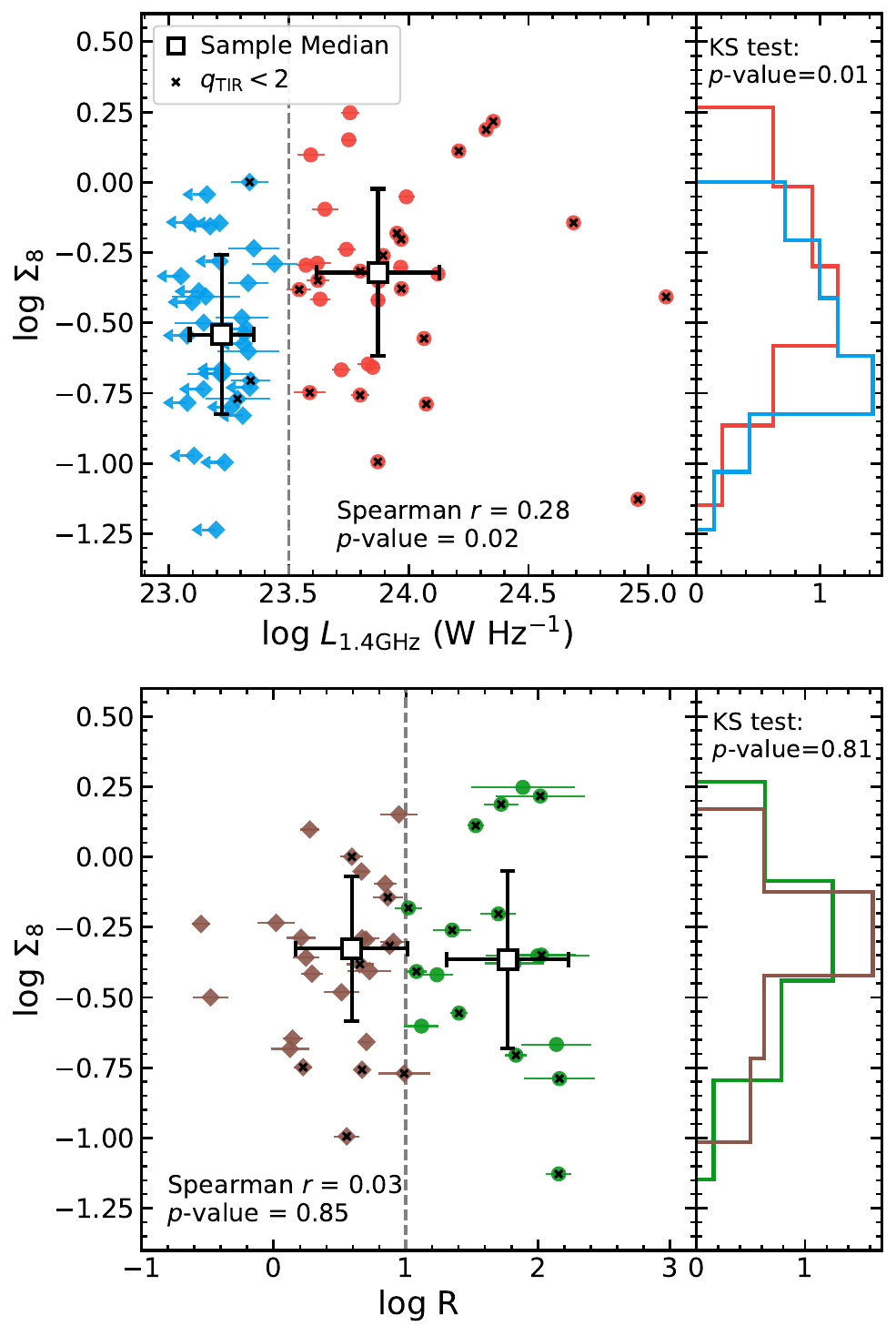}
\caption{Environmental dependence of radio luminosity and radio loudness. The upper panel shows a weak positive correlation between environmental density and radio luminosity, with AGNs separated at $\log L_{1.4{\rm GHz}} ({\rm W\ Hz^{-1}})=23.5$. The lower panel shows no correlation for radio loudness, with AGNs separated at $\log R=1$. 
The open squares show the median values with 1$\sigma$ error bars for separated samples. AGNs with $q_{\rm TIR}<2$ are marked by black crosses. 
The Spearman test $r$ and $p$-value are given in each panel.
The environmental density distributions of the separated samples are shown in each right subpanel with their KS test $p$-value. 
 \label{fig:radio}}
\end{figure}

\begin{figure*}[t]
\epsscale{1. }
\plotone{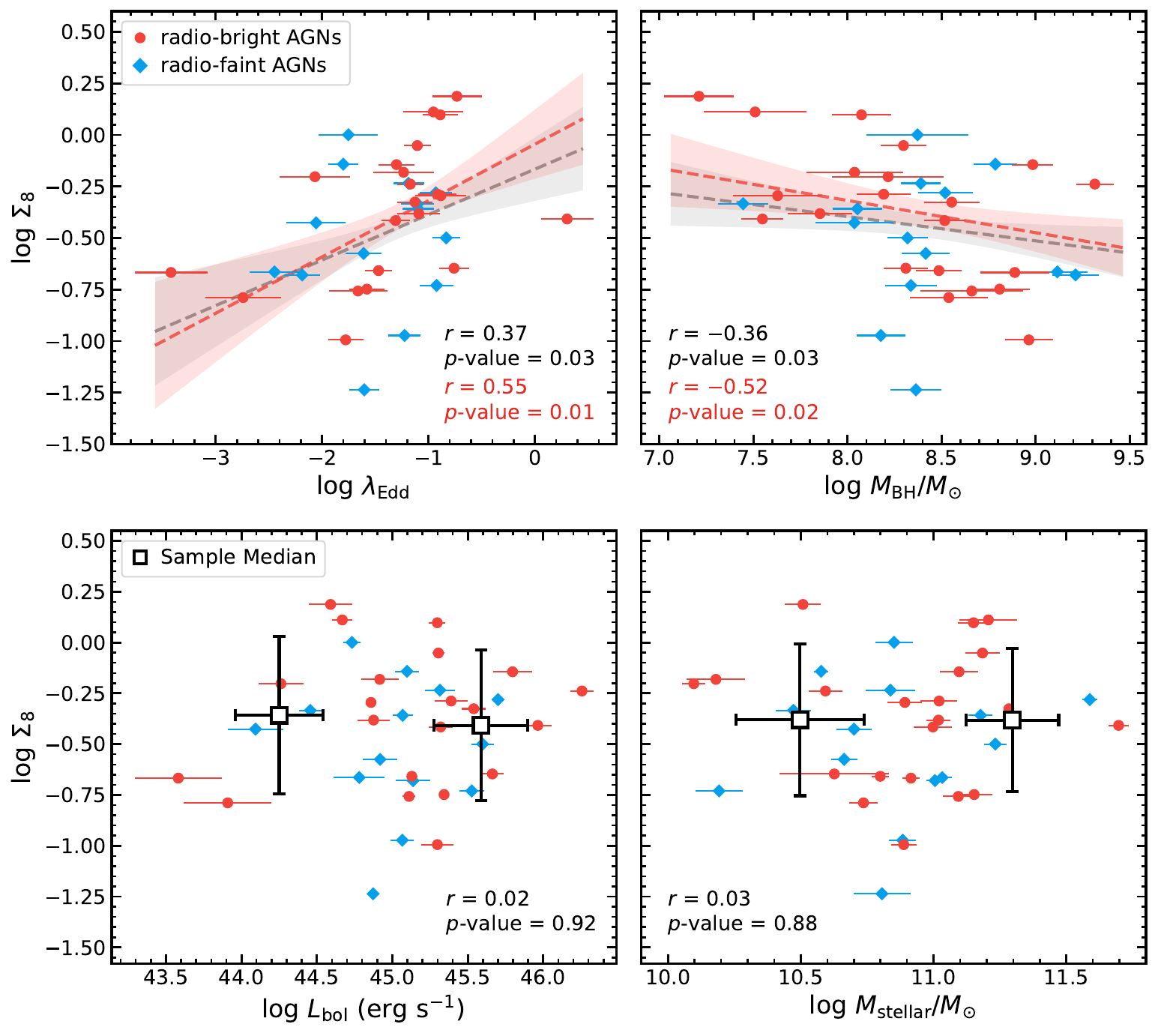}
\caption{Environmental dependence of AGN properties. In each panel, radio-bright and radio-faint AGNs are shown as red circles and blue diamonds, respectively. The upper panels show environmental dependence on $\lambda_{\rm Edd}$ and $M_{\rm BH}$. The dashed lines with shaded 1$\sigma$ uncertainty region indicate the best-fit linear relations for the full AGN sample (in grey) and the radio-bright AGN sample (in red). The lower panels show no environmental dependence on the AGN $L_{\rm bol}$ or host-galaxy $M_{\rm stellar}$, with the open squares showing the binned sample median with 1$\sigma$ error bars. 
The Spearman $r$ and $p$-value are given in each panel for the full AGN sample (in black) and radio-bright AGN sample (in red).
\label{fig:depend}}
\end{figure*}

We find a weak positive correlation between environmental density and radio luminosity, with a Spearman test coefficient $r$ = 0.28 and $p$-value = 0.02. This trend is shown in the upper panel of Figure~\ref{fig:radio}. The radio-bright and radio-faint AGN samples exhibit significantly different environmental density distributions, with a KS test $p$-value = 0.01. Meanwhile, environmental density shows no dependence on radio loudness. As shown in the lower panel of Figure~\ref{fig:radio}, radio-loud ($\log R > 1$) and radio-quiet ($\log R < 1$) AGNs have statistically similar environmental density distributions.
The environmental densities also do not depend on $M_{\mathrm{stellar}}$ within the AGN sample, and therefore are not biased by massive hosts. 
These results suggest that denser environments are associated with enhanced radio power in our probed radio luminosity range, while the radio-emission mode of AGNs remains unchanged. We also investigate the environmental dependence on X-ray luminosity and $q_{\rm TIR}$, finding no significant correlations.

We notice a potential decrease in environmental density toward the high-radio-luminosity end. This trend is consistent with \cite{malavasi2015}, which find an environmental distinction between low- and high-power radio AGNs with a threshold of $\log L_{1.4{\rm GHz}}({\rm W\ Hz^{-1}})=24.5$. Given the small number of high-power radio AGNs in our sample, we are not able to reach a robust conclusion. Future larger-area observations are needed.

The AGN-dominated radio sources with $q_{\rm TIR}<2$ are not significantly separated from the rest of the sample, although they exhibit a larger scatter in environmental density. We mark these sources in Figure~\ref{fig:radio}.
To test whether the observed environmental dependence of radio luminosity could be driven solely by star formation, we analyze a sample of Pa$\alpha$ radio galaxies with $q_{\rm TIR}>2.2$, excluding our selected AGNs. The sample construction and environmental measurements are described in Appendix~\ref{sec:appendixB}. We find no correlation between radio luminosity and environmental density, nor any density excess for radio-bright galaxies with $\log L_{1.4{\rm GHz}}({\rm W\ Hz^{-1}})>23.5$.
Radio galaxies powered by star formation cannot reproduce the environmental dependence observed in the AGN sample. Although the observed radio emission generally originates from both AGN activity and host-galaxy star formation, the enhanced radio luminosities found in denser environments require an AGN-related contribution. 

We find evidence for correlations between environmental density and black hole properties. Using the broad-line Pa$\alpha$ AGN sample, we identify a positive correlation between environmental density and $\lambda_{\rm Edd}$, and a slightly negative correlation with $M_{\rm BH}$. Both trends become stronger when restricting to the radio-bright AGN sample. The corresponding Spearman test $r$ and $p$-value are shown in Figure~\ref{fig:depend}.
We adopt an additional uncertainty of 0.1 dex for all $M_{\rm BH}$ measurements. For the super-Eddington source ID7741 ($\lambda_{\rm Edd}>1$), we further add a 0.2 dex uncertainty to $M_{\rm BH}$ because a nearby emission contaminates its broad Pa$\alpha$ line. We then perform linear least-squares fittings. The best-fit relations of $\log \Sigma_8 = a\times\log \lambda_{\rm Edd}+b$ and $\log \Sigma_8 = c\times\log(M_{\rm BH}/M_\odot)+d$ are $(a,b,c,d)=(0.22\pm0.11,-0.17\pm0.15,-0.12\pm0.11,0.54\pm0.90)$ for the full broad-line AGN sample and $(0.27\pm0.13,-0.04\pm0.17,-0.16\pm0.12,0.93\pm1.0)$ for the radio-bright broad-line AGN sample.

We find no significant environmental dependence on AGN $L_{\rm bol}$ or host-galaxy $M_{\rm stellar}$ in this sample, suggesting that the observed trends are not driven by luminosity- or mass-related biases.
Moreover, the grism spectroscopy used to identify Pa$\alpha$ emitters is unlikely to introduce selection biases into the broad-line Pa$\alpha$ AGN environments. Therefore, our results suggest enhanced AGN accretion in the dense environments traced by Pa$\alpha$ emitters.

\subsection{Luminosity Functions and Overdensities} \label{sec:lfs}

We construct LFs for the Pa$\alpha$ emitters in all AGN and radio-bright AGN environments, which are hereafter referred to environmental LFs.
We also construct a LF of overall field Pa$\alpha$ emitters for comparison. The construction of the selection function is described in Section~\ref{sec:emitters}. 
We first compute binned LFs using the standard $1/V_a$ method \citep{avni1980}, accounting for the selection completeness. The uncertainty of the binned LF is the quadrature sum of the Poisson error and 15\% cosmic variance in the COSMOS-3D field \citep{moster2011a}. For the environmental LFs, the effective survey volume is the sum of all individual AGN environmental regions used in the surface density calculation.
We further parametrize the LFs using a Schechter function \citep{schechter1976},
\begin{equation}
\Phi(L)dL = \Phi_* \left(\frac{L}{L_*}\right)^{\alpha} \exp(-L/L_*) \frac{dL}{L_*}.
\end{equation}
We perform a Markov Chain Monte Carlo (MCMC) analysis using the Python code {\tt emcee} \citep{foreman-mackey2013} to determine the best-fit parameters $\Phi_*$, $\alpha$, and $L_*$. The resulting LFs are shown in Figure~\ref{fig:lfs}, with the corresponding values presented in Appendix~\ref{sec:appendixC}.

The environmental LFs are systematically higher than the field Pa$\alpha$ LF, particularly for the radio-bright AGN sample. At $\log L_{\mathrm{Pa\alpha}} (\mathrm{erg\ s^{-1}})\sim 41.5-42.0$, the number density of Pa$\alpha$ emitters in radio-bright AGN environments is $\sim2$–$3$ times higher than in the field.
We compute the overdensity as $\delta = N/\bar{N} - 1$, where $N$ is the incompleteness-corrected count within the aperture defined by the eighth-nearest neighbor, and $\bar{N}$ is the expected number derived from the field Pa$\alpha$ LF within the same aperture. We adopt a luminosity cut of $\log L_{\mathrm{Pa\alpha}}\ (\mathrm{erg\ s^{-1}}) \geq 40.8$.
We find that 66\% of the full AGN sample and 74\% of the radio-bright AGN sample reside in overdense regions ($\delta > 0$). This indicates that radio-bright AGNs are efficient tracers of large-scale overdensities. 
We note that this overdensity is measured relative to the overall Pa$\alpha$ emitter population, instead of the control samples used in Section~\ref{sec:envir}.

We find a flattening trend at the faint end of the environmental LFs, with the faint-end slope $\alpha$ becoming shallower from the field to the AGN environment, and further to the radio-bright AGN environment. This trend is consistent with literature results on H$\alpha$ LFs and galaxy stellar mass functions in overdense regions \citep[e.g.,][]{vanderburg2020,forrest2024,lin2026b}.
The shallower faint-end slope may reflect accelerated galaxy growth and environmental processes. Simulations of galaxy evolution find early formation of massive galaxies and rapid merging and quenching activity in proto-cluster environments compared to field galaxies, driven by the high halo density and top-heavy halo mass functions \citep{muldrew2018}.

\begin{figure}[t]
\epsscale{1.2}
\plotone{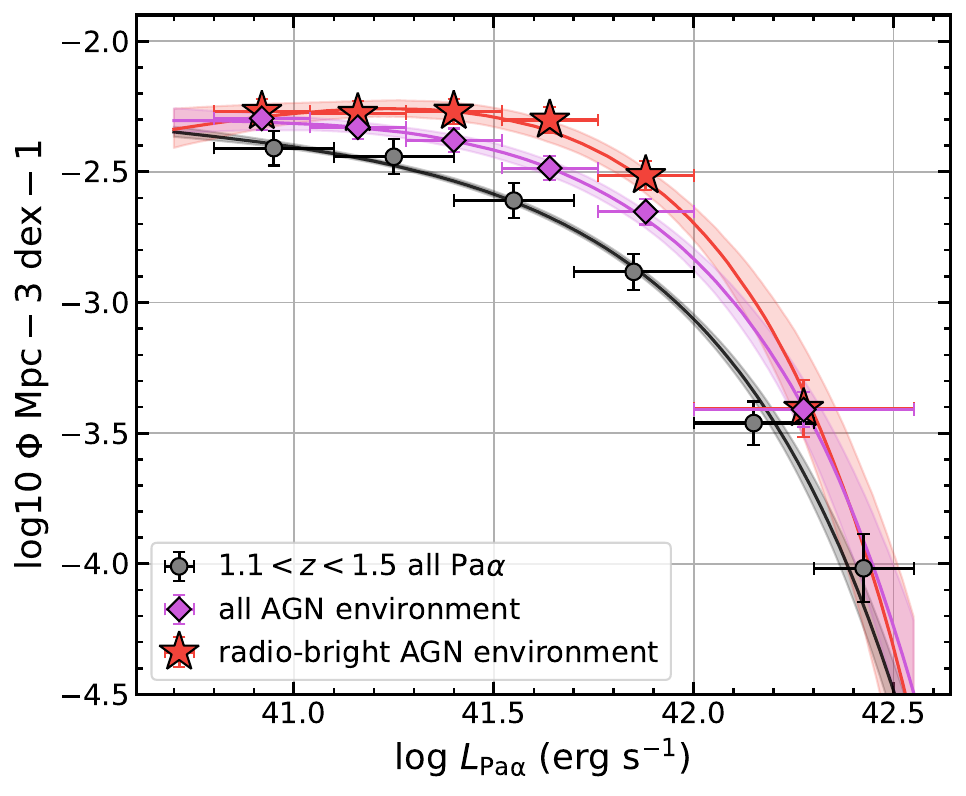}
\caption{Pa$\alpha$ LFs at $z\sim 1.1-1.5$. 
Three binned LFs for Pa$\alpha$ emitters in field, all AGN environments, and radio-bright AGN environments are shown as dark grey circles, purple diamonds, and red stars, respectively.
Their horizontal bars indicate the luminosity ranges covered by the bins, and the vertical bars indicates the 1$\sigma$ errors. The corresponding lines with the shaded regions represent best-fit LFs with their 1$\sigma$ uncertainty regions. 
 \label{fig:lfs}}
\end{figure}

\section{Summary and Implication} \label{sec:sum}

We have statistically investigated the large-scale environments of 70 AGNs using a highly complete Pa$\alpha$ emitter sample in the COSMOS-3D field at $z=1.1$–$1.5$. We constructed $M_{\rm stellar}$- and SFR-matched control samples and measured the eighth-nearest-neighbor surface density of Pa$\alpha$ emitters around both AGNs and control galaxies. This environmental density probes Mpc-scale environments. 
We confirm that the environmental density distribution of radio-bright AGNs is statistically higher than that of their control galaxies. We further find a weak positive correlation between environmental density and radio luminosity, together with a positive correlation with $\lambda_{\rm Edd}$ and a weak negative correlation with $M_{\rm BH}$, particularly for the radio-bright AGN sample.
We also construct Pa$\alpha$ LFs in AGN environments and find that they exhibit shallower faint-end slopes and systematically higher number densities than the field Pa$\alpha$ LF. 

Since Pa$\alpha$ emitters trace actively star-forming galaxies, regions with higher Pa$\alpha$ emitter densities are likely associated with more abundant cold-gas reservoirs.
The environmental dependence of $\lambda_{\rm Edd}$ and $M_{\rm BH}$ provides direct evidence for enhanced AGN accretion and rapid SMBH growth in gas-rich environments. The shallower faint-end slopes of the environmental LFs further support the accelerated galaxy evolution.
The weak positive environmental dependence of radio luminosity suggests that radio emission is moderately enhanced in denser environments. 
Other processes related to the central engine, such as black hole spin and host-galaxy gas dynamics, are also expected to regulate the radio power of AGNs and introduce scatter to the environmental dependence.
Moreover, the observed radio emission contains contributions from both AGN activity and host-galaxy star formation. In addition to producing radio emission, AGN activity may indirectly boost the total radio output by enhancing star formation through positive feedback  \citep[e.g.,][]{kalfountzou2014,kalfountzou2017}.

Overall, our results support a scenario in which large-scale supply of cold gas could regulate both black-hole growth and galaxy evolution at cosmic noon. High-quality JWST WFSS surveys have opened a new era for AGN environmental studies. The upcoming Roman Space Telescope will provide environmental measurements over a larger cosmic volume. 
In the future, growing spectroscopic datasets will enable more comprehensive environmental investigations for diverse AGN populations at different redshifts.

\begin{acknowledgments}
We acknowledge support from the National Science Foundation of China (12225301). J.-T.S. acknowledges funding by the Deutsche Forschungsgemeinschaft (DFG, German Research Foundation) - Project number 518006966 - and support by the DFG under Germany’s Excellence Strategy – EXC 2121 ``Quantum Universe'' – 390833306.
\end{acknowledgments}

\facilities{JWST}
\software{{\tt\string Astropy} \citep{astropy}, {\tt\string Numpy} \citep{numpy}, {\tt\string Scipy} \citep{scipy}, {\tt\string Matplotlib} \citep{matplotlib}, {\tt\string emcee} \citep{foreman-mackey2013}.}

\bibliography{ms}
\bibliographystyle{aasjournalv7}

\appendix

\section{Selection Completeness of the line emitter sample}\label{sec:appendixA}
\begin{figure}[t]
\epsscale{0.7}
\plotone{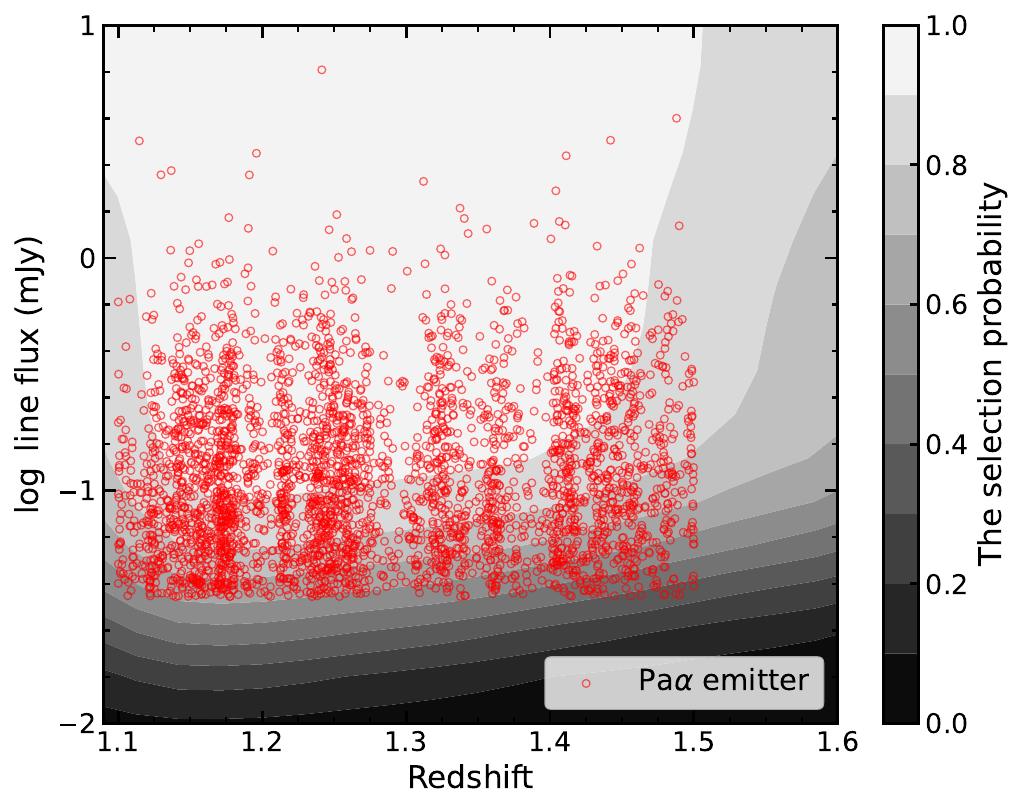}
\caption{Selection probabilities as functions of line flux and redshift. The contours are selection probabilities from 0 to 1 with an interval of 0.1, as shown in the color bars. The open red circles indicate the locations of 3,419 Pa$\alpha$ emitters in the selection function. \label{fig:completeness}}
\end{figure}

We perform simulations to derive the Pa$\alpha$ selection completeness.
In the simulation, we first extract local 2D spectra on a uniform grid of mock positions spaced by $10\arcsec$ across the field. For each extracted spectrum, we generate 20 mock realizations by injecting emission-line source models. The continuum is modeled as a power law ($f_{\lambda}\propto\lambda^{\alpha}$), normalized to reproduce the observed continuum magnitude distribution.
To account for the extended morphologies of Pa$\alpha$ emitters, we use real F444W images as 2D line templates. 
We isolate the target source, resample the image to the grism pixel scale, and scale it following the observed line flux distribution. We then add the corresponding Poisson noise from the injected sources to the inverse-variance maps. We apply the same line-detection procedure as used for the real data, and retain detections whose recovered line centers lie within two instrumental full width at half maximum (FWHM) of the input wavelength. The resulting selection function is computed as the fraction of recovered simulated lines in each bin of line flux and redshift. Figure \ref{fig:completeness} shows the selection function, with Pa$\alpha$ emitters shown as red open circles.

\section{Radio Galaxy Environments}\label{sec:appendixB}
\begin{figure}[t]
\epsscale{0.7}
\plotone{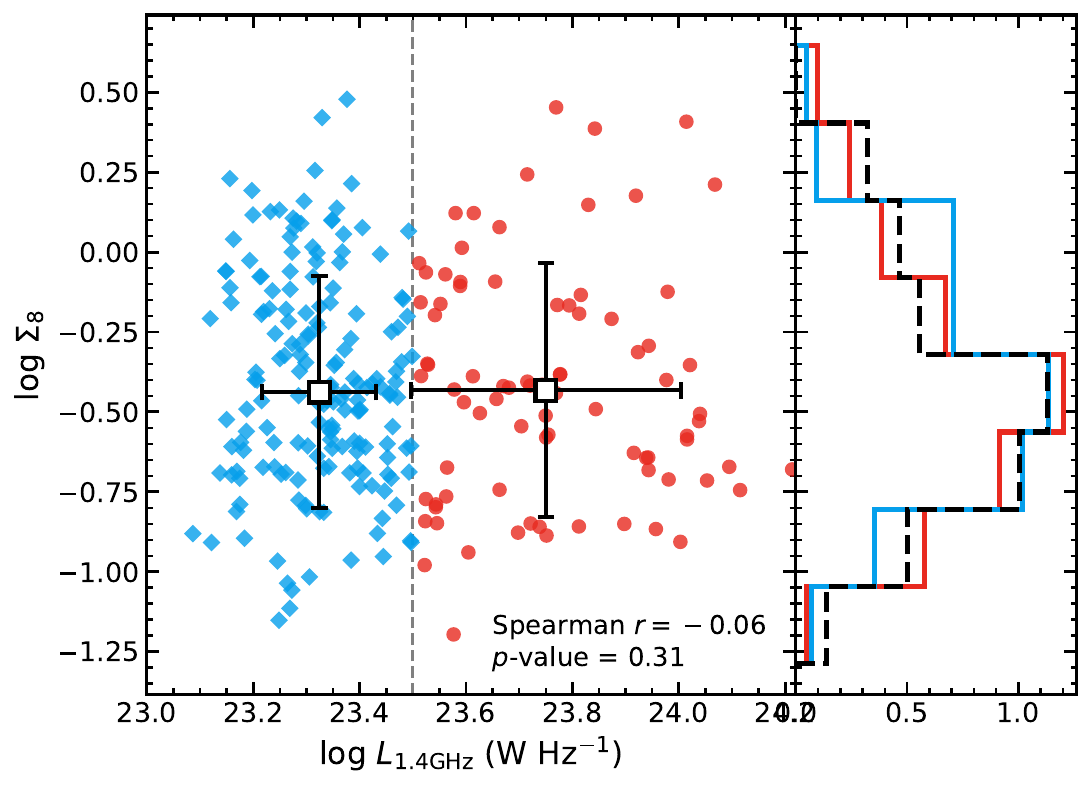}
\caption{Environmental dependence of radio luminosity for radio galaxies. Similar to the upper panel of Figure~\ref{fig:radio}. The typical uncertainty in radio luminosity is 0.08 dex. In the right panel, the black dashed line shows the median environmental densities of the radio-bright control sample. The radio-bright galaxies and their controls have indistinguishable density distributions with KS test $p$-value = 0.81.
 \label{fig:rg_appendix}}
\end{figure}

We construct a sample of Pa$\alpha$ radio galaxies and investigate their environments. We cross-match our Pa$\alpha$ emitter catalog with the radio catalogs \citep{smolcic2017,jin2018}, excluding our selected AGNs. We calculate the radio luminosities and $q_{\rm TIR}$ values for the matched radio galaxies following the same procedure in Section~\ref{sec:paa_agns}.
We then select 268 radio galaxies with $q_{\rm TIR}>2.2$, indicating that their radio emission is predominantly powered by star formation. We measure the environmental densities of these galaxies and find no significant dependence on radio luminosity. We further select 92 radio-bright galaxies with $\log L_{1.4{\rm GHz}}({\rm W\ Hz^{-1}})>23.5$ and construct $M_{\rm stellar}$- and SFR-matched control samples. No statistically significant difference is found between the environmental density distributions of the radio-bright and control samples. The corresponding results are presented in Figure~\ref{fig:rg_appendix}.

\section{Parameters of the luminosity functions}\label{sec:appendixC}
We present the binned LFs in Table \ref{tab:binned_LF} and the parametrized LFs in Table \ref{tab:LF}. The construction of those LFs is in Section \ref{sec:lfs}.

\begin{deluxetable}{cccc}[t]
\tablecaption{Binned Luminosity Function\label{tab:binned_LF}}
\setlength{\tabcolsep}{10pt}
\tablehead{
\colhead{$\log L_{\mathrm{P}\alpha}$} &
\colhead{$\Delta \log L$} &
\colhead{$\log \Phi^{a}$} &
\colhead{$N$}
}
\startdata
\multicolumn{4}{c}{all $\mathrm{Pa}\alpha$ at $1.1<z<1.5$} \\
\hline
40.95 & 0.30 & $-2.41\pm$0.07 & 1100 \\
41.25 & 0.30 & $-2.44\pm$0.07 & 982 \\
41.55 & 0.30 & $-2.61\pm$0.07 & 627 \\
41.85 & 0.30 & $-2.88\pm$0.07 & 309 \\
42.15 & 0.30 & $-3.45\pm$0.08 & 75 \\
42.42 & 0.25 & $-4.02\pm$0.13 & 15 \\
\hline
\multicolumn{4}{c}{all AGN environmental $\mathrm{Pa}\alpha$} \\
\hline
40.92 & 0.24 & $-2.27\pm$0.09 & 60 \\
41.16 & 0.24 & $-2.28\pm$0.09 & 57 \\
41.40 & 0.24 & $-2.27\pm$0.09 & 56 \\
41.64 & 0.24 & $-2.30\pm$0.09 & 49 \\
41.88 & 0.24 & $-2.51\pm$0.10 & 28 \\
42.27 & 0.55 & $-3.41\pm$0.18 & 7 \\
\hline
\multicolumn{4}{c}{radio-bright AGN environmental $\mathrm{Pa}\alpha$} \\
\hline
40.92 & 0.24 & $-2.30\pm$0.07 & 140 \\
41.16 & 0.24 & $-2.33\pm$0.08 & 125 \\
41.40 & 0.24 & $-2.38\pm$0.08 & 107 \\
41.64 & 0.24 & $-2.49\pm$0.08 & 79 \\
41.88 & 0.24 & $-2.65\pm$0.09 & 50 \\
42.27 & 0.55 & $-3.41\pm$0.12 & 17 \\
\enddata
\tablenotetext{a}{Luminosity function in units of $\mathrm{Mpc^{-3}\,dex^{-1}}$.}
\end{deluxetable}

\begin{deluxetable}{cccc}
\tablecaption{Parametrized Luminosity Function\label{tab:LF}}
\setlength{\tabcolsep}{4pt}
\tablehead{ 
\colhead{Sample} &
\colhead{$\log L^*$} &
\colhead{$\log \Phi^{*}$} &
\colhead{$\alpha$} 
}
\startdata
all $\mathrm{Pa}\alpha$ at $1.1<z<1.5$ & $41.83_{-0.03}^{+0.04}$ & $-2.77_{-0.04}^{+0.04}$ & $-1.08_{-0.05}^{+0.05}$ \\
all AGN environmental $\mathrm{Pa}\alpha$ & $41.82_{-0.07}^{+0.08}$ & $-2.54_{-0.09}^{+0.08}$ & $-0.92_{-0.11}^{+0.11}$ \\
radio-bright AGN environmental $\mathrm{Pa}\alpha$ & $41.72_{-0.09}^{+0.10}$ & $-2.32_{-0.10}^{+0.08}$ & $-0.67_{-0.16}^{+0.17}$ \\
\enddata
\end{deluxetable}

\end{document}